\documentstyle[12pt]{article}
\begin{document}
\title{\bf Action Correlations in Integrable Systems}
\author{E. Bogomolny\\
\\
Laboratoire de Physique Th\'eorique\\
et\\
Mod\`eles Statistiques\footnote{Unit\'e Mixte de Recherche de l'Universit\'e
  Paris XI et du CNRS (UMR 8626)}\\
\\
Universit\'e Paris - Sud\\
91405 Orsay Cedex, France}
\maketitle
\begin{abstract}
In many problems of quantum chaos the calculation of sums of products 
of periodic orbit contributions is required. A general method of computation
of these sums is proposed for generic integrable models where the summation
over periodic orbits is reduced to the summation over integer vectors
uniquely associated with periodic orbits. It is demonstrated that in
multiple sums over such integer vectors there exist hidden saddle points
which permit explicit evaluation of these sums. Saddle point manifolds consist
of periodic orbits vectors which are almost mutually parallel. Different
problems has been treated by this saddle point method, e.g. Berry's
bootstrap relations, mean values of Green function products etc. In
particular, it is obtained that suitably defined 2-point correlation
form-factor for periodic orbit actions in generic integrable models is
proportional to quantum density of states and has peaks at quantum
eigenenergies.
\end{abstract}

\newpage

\section{Introduction}
The trace formulas are the main instrument in the investigation of relations 
between quantum and classical properties of a given system (see e.g. 
\cite{1}). These formulas connect the quantum density of states, $d(E)$,
in the semi-classical limit with a sum over classical periodic orbits (po)
\cite{1}-\cite{4}
\begin{equation}
d(E)=\bar{d}(E)+\sum_{po}A_p
\exp(i\frac{S_p(E)}{\hbar}),
\label{1}
\end{equation}
where $\bar{d}(E)$ is the mean density, $S_p(E)$ is the classical action along 
a periodic orbit, and $A_p$ is a  pre-factor build from classical quantities.

There is two main applications of these formulas. First, one tries to use 
them to compute smoothed density of states (or even approximate positions of 
energy levels \cite{6}). Second, one uses trace formulas to understand 
statistical properties of energy levels \cite{7}. In the latter approach 
one starts with a formal expression of $n$-point correlation function,
$R_n(\epsilon_1, \epsilon_2,\ldots, \epsilon_n)$, as a mean value of 
product of $n$ density of states, $d(E)$, at different points
\begin{equation}
R_n(\epsilon_1, \epsilon_2, \ldots, \epsilon_n)=<d(E+\epsilon_1)
d(E+\epsilon_2)\cdots d(E+\epsilon_n)>,
\end{equation}
where the brackets denote an average over an interval of energy $\Delta E$
which is classically small but includes a large number of quantum levels.
Substituting in this formula the semi-classical expression (\ref{1}) for 
the density of states one obtains a semi-classical approximation for 
correlation functions.

The main difficulty of such method is the calculation of mean values of 
products of contributions of different periodic orbits
\begin{equation}
<\sum_{p_1,p_2,\ldots}\exp (i\frac{1}{\hbar}\sum_j e_j S_{p_j}(E))>,
\label{3}
\end{equation}
where $e_j=\pm 1$. When all $e_j$ are of the same sign this average equals 
zero but if signs of $e_j$ are different the calculations are far from being
clear. 

The simplest approximation (called the diagonal approximation) consists of 
taking  into account only terms with exactly the same action
\begin{equation}
<\exp i\frac{1}{\hbar}(S_{p_1}-S_{p_2})>=\left\{
\begin{array}{l} 1,\;\mbox{if } S_{p_1}=S_{p_2}\\
              0, \;\mbox{if } S_{p_1}\neq S_{p_2}
\end{array} \right . .
\end{equation}
Berry \cite{7} showed that for integrable systems this approximation 
gives the correct 2-point correlation form-factor but for chaotic systems
the validity of diagonal approximation is limited to short - time behavior 
of the form-factor. For more complicated quantities (even for integrable
systems) this approach is insufficient and more refined methods are needed 
\cite{8}-\cite{10}.
 
The purpose of this paper is to develop a method which permits explicit 
computations of the expressions (\ref{3}) for generic {\bf integrable} 
systems. We shall show that in these expressions there exist hidden 
saddle points which give dominant contribution in the semi-classical limit.

The usual trace formulas can be applied only to quantum mechanical quantities
which (like in Eq.~(\ref{1})) are expressed through {\bf one} Green function.
The method discussed in this paper permits to construct a new type of trace
formulas which express a product of such quantities for integrable systems
through a sum over classical periodic orbits.  

The plan of the paper is the following. In Section 2  for completeness the 
derivation of the trace formula for a rectangular billiard is presented.
This is a simple prototype of  integrable models where all formulas
are transparent without unnecessary complications.
In Section 3 on the example of Berry's bootstrap 
we demonstrate the existence  and importance of quantities which include sums like 
in Eq.~(\ref{3}) and whose semi-classical calculation is the main topic of 
the rest of the paper.  The saddle point method  for oscillatory and smooth 
terms is discussed in Sections 4 and 5 respectively. In is shown that the
dominant saddle point configurations are build from periodic orbits whose 
integer vectors are almost mutually parallel. In Section 6 
the calculation of mean values of products of different powers of retarded
and advanced Green functions is presented. In this Section we also 
discuss the calculation of next-to-leading terms by the saddle point
method. The correlation functions for classical actions are  derived in Section 7.
An interesting consequence of the discussed formalism is the fact that
(properly defined) 2-point correlation form-factor for actions of classical 
periodic orbits is proportional to quantum density of states. 
The generalization of the proposed method for generic integrable models
is performed in Section 8.

\section{Trace formula for rectangular billiard}
Let us consider a plane rectangular billiard with sides $a$ and $b$ with 
(for simplicity) periodic boundary conditions. The functions 
\begin{equation}
\Psi_{m,n}(x,y)=\frac{1}{\sqrt{ab}}\exp (\frac{2\pi i}{a}nx+
\frac{2\pi i}{b}my),
\end{equation}
with integers $m,n=0,\pm 1,\pm 2,\ldots $ are periodic solutions of the 
(Schr\"odinger) equation
\begin{equation}
(E_{mn}+\Delta )\Psi_{m,n}(x,y)=0
\end{equation}
with energy 
\begin{equation}
E_{mn}=(\frac{2\pi}{a})^2n^2+(\frac{2\pi}{b})^2m^2.
\end{equation}
The quantum density of these states
\begin{equation}
d(E)=\sum_{m,n=-\infty}^{\infty}\delta(E-E_{mn})
\end{equation}
may be rewritten by the Poisson summation formula in the following way
\begin{equation}
d(E)=\sum_{M,N=-\infty}^{\infty}\int e^{2\pi i(Mm + Nn)}
     \delta(E-(\frac{2\pi}{a})^2n^2-(\frac{2\pi}{b})^2m^2)dndm.
\end{equation}
The substitution $n=ar\cos \theta /(2\pi)$ and $m=br\sin \theta /(2\pi)$                           
after simple algebra gives
\begin{equation}
d(E)=\frac{ab}{4\pi}\sum_{M,N=-\infty}^{\infty}J_0(kL_{MN}),
\label{11}
\end{equation}
where $E=k^2$, 
\begin{equation}
L_{MN}=\sqrt{(Na)^2+(Mb)^2},
\label{12}
\end{equation}
and
\begin{equation}
J_0(x)=\frac{1}{2\pi}\int_0^{2\pi}e^{ix\cos \theta}d\theta
\end{equation}
is the Bessel function of zero order. It is evident that $L_{MN}$ is the 
length of a classical periodic orbit or, more strictly, the length of a
classical trajectory on a resonant (= periodic) torus. In integrable models    
almost all periodic orbits belong to resonant tori and we shall use
these two notions on the equal footing.

As  most of the terms in both parts of Eq.~(\ref{11}) are 4 times degenerated
it is convenient to define the density of non-degenerate states by dividing 
both parts of this equation by 4
\begin{equation}
d(E)=\bar{d}(E)+d^{osc}(E).
\label{14}
\end{equation}
Here
\begin{equation}
\bar{d}(E)=\frac{A}{16\pi},
\end{equation}
where $A=ab$ is the area of the rectangle and 
\begin{equation}
d^{osc}(E)=\frac{A}{4\pi}\sum_{L_p} e_pJ_0(kL_p),
\end{equation}
where the sum is taken over all periodic orbits defined by two integers
$M$ and $N$ whose lengths $L_p$ are given by Eq.~(\ref{12}). $e_p=1$ if both
$M$ and $N$ are nonzero and $e_p=1/2$ otherwise (the term with $M=N=0$ gives
$\bar{d}(E)$).

In semi-classical limit $k\rightarrow \infty$ and 
\begin{equation}
d^{osc}(E)=\frac{A}{2\pi}\sum_{L_p}\frac{e_p}{\sqrt{8\pi kL_p}}
e^{ikL_p-i\pi/4}+c.c.
\label{17}
\end{equation}
Eqs.~(\ref{14})-(\ref{17}) define the trace formulas for a rectangular 
billiard (with periodic boundary conditions). Trace formula for general 
integrable systems have  similar form and will be discussed in Section 8.

\section{Berry's bootstrap}
 
The above trace formula for rectangular billiard is the simplest example 
of trace formulas for integrable systems. We know explicitly everything. 
The periodic orbit lengths are given by simple expression (\ref{12}) and 
all semi-classical corrections to Eq.~(\ref{17}) (coming from asymptotic 
expansion of the Bessel function) are easy to take into account.

But even in this case quantities like those in Eq.~(\ref{3}) is difficult 
to compute.
The necessity of such calculation is clearly seen e.g. from Berry's remark
\cite{7} that for any quantum system with non-degenerated discrete spectrum
the square of the density of states should be proportional to the density 
itself. 

This fact can be demonstrated as follows. The density of states at real 
values of energy is usually defined as the limit $\epsilon \rightarrow 0$
of the sum over all energy eigenvalues $e_n$
\begin{equation}
d_{\epsilon}(E)=\frac{1}{2\pi i}\sum_n(\frac{1}{E-e_n-i\epsilon}-
\frac{1}{E-e_n+i\epsilon})=
\frac{\epsilon}{\pi}\sum_n\frac{1}{(E-e_n)^2+\epsilon^2}.
\label{18}
\end{equation}
In the semi-classical trace formula a finite value of $\epsilon$ corresponds
to a regularization of divergent sums by adding (e.g. in Eq.~(\ref{17}))
a small imaginary part to the energy $E\rightarrow E+i\epsilon$.

As it is evident from Eq.~(\ref{18}) at finite (but small) values of 
$\epsilon$ the function $d_{\epsilon}(E)$ has peaks near each energy 
eigenvalues, the pear width being proportional to $\epsilon$. Now let us 
compute $d_{\epsilon}^2(E)$ assuming that all eigenvalues are 
non-degenerated. It is clear that in the limit $\epsilon \rightarrow 0$
\begin{equation}
d_{\epsilon}^2(E)=\frac{\epsilon^2}{\pi^2}
\sum_n\frac{1}{((E-e_n)^2+\epsilon^2)^2},
\end{equation}
and cross terms with different $E_n$ will give negligible contributions
at small $\epsilon$. 

Because
\begin{equation}
\int_{-\infty}^{\infty}\frac{dx}{(x^2+\epsilon^2)^2}=\frac{\pi}{2\epsilon^3},
\end{equation}
one gets that
\begin{equation}
\lim_{\epsilon \rightarrow 0}\frac{\epsilon^3}{(x^2+\epsilon^2)^2}=
\frac{\pi}{2}\delta (x),
\end{equation}
and therefore
\begin{equation}
\lim_{\epsilon \rightarrow 0} 2\pi \epsilon d_{\epsilon}^2(E)=d(E).
\label{22}
\end{equation}
Similarly by computing higher powers of $d_{\epsilon}(E)$ one obtains that
\begin{equation}
\lim_{\epsilon \rightarrow 0} \pi^{n-1}\frac{(2n-2)!!}{(2n-3)!!}
   \epsilon^{n-1} d_{\epsilon}^n(E)=d(E).
\label{23}
\end{equation}
These bootstrap relations reflect the analytical structure of the 
density of states, namely that it should have $\delta$-function 
singularities at real values of energy with unit residues. Therefore these 
type of relations is very general and our first purpose is to find a method 
which will permit to prove these relations for integrable systems starting 
from the semi-classical trace formula (\ref{17}).

\section{Saddle points for oscillatory terms} 
Let us begin with the case $n=2$. The density of states may be written as a sum of 
two terms $d_{\epsilon}(E)=d_{+\epsilon}(E)+d_{-\epsilon}(E)$ where $\pm$ 
corresponds to the sign of the exponent in two terms in Eq.~(\ref{17})
\begin{equation}
d_{+\epsilon}(E)=\frac{A}{2\pi}\sum_{L_p}\frac{e_p}{\sqrt{8\pi kL_p}}
e^{ikL_p-i\pi/4-\epsilon L_p/2k},
\end{equation}
and $d_{-\epsilon}(E)=d_{+\epsilon}^{*}(E)$. Of course, the contribution of 
$\bar{d}(E)$ to the left-hand side of bootstrap relations is negligible at
small $\epsilon$. Therefore 
$d_{\epsilon}^2(E)=d_{+\epsilon}^2(E)+2d_{+\epsilon}(E)d_{-\epsilon}(E)+
d_{-\epsilon}^2(E)$. The periodic orbit expansion of  
$d_{\pm \epsilon}^2(E)$ contains only terms with the same sign of the actions
in the exponent and for any {\bf fixed} total action it includes only
a finite number of terms, and in the limit $\epsilon \rightarrow 0$ 
their contributions are negligible. On the contrary the periodic orbit 
expansion of $d_{+\epsilon}(E)d_{-\epsilon}(E)$ for any finite total action
contains infinite number of terms and it is this term which can give 
contribution in the limit of small $\epsilon$. Therefore the following relation
should be true
\begin{eqnarray}
&&\lim_{\epsilon \rightarrow 0} 2\pi \epsilon 
\frac{A^2}{2\pi^2}\sum_{p_1,p_2}\frac{e_{p_1}e_{p_2}}{8\pi k\sqrt{l_1l_2}}
e^{ik(l_1-l_2)-(l_1+l_2)\epsilon/2k}\nonumber\\
&=&\bar{d}(E)+\frac{A}{2\pi}\sum_{p}(\frac{e_p}{\sqrt{8\pi kl_p}}
e^{ikl_p-i\pi/4} +c.c.).
\end{eqnarray}
Here in the left-hand side $l_{1,2}$ are the lengths of periodic orbits 
$p_1$ and $p_2$ and the sum is taken over all such orbits.

The smooth part (=mean value) of this expression is easy to compute. It was
shown in \cite{7} that diagonal approximation is applicable and the mean 
value of the level density is expressed through the sum over periodic orbits
as follows
\begin{equation}
\lim_{\epsilon \rightarrow 0} 2\pi \epsilon
\frac{A^2}{2\pi^2}\sum_{p}\frac{e^{-\epsilon l/k}}{8\pi k l}=\bar{d}(E).
\label{25}
\end{equation}
The density of periodic orbits can  easily be calculated from Eq.~(\ref{12}).
If $N(l)$ is the number of periodic orbits with the lengths less than $l$
then for large $l$
\begin{equation}
N(l)=\frac{\pi l^2}{4A},
\label{26}
\end{equation}
where $A=ab$ is the area of the rectangle and the factor $1/4$ comes because 
we consider only orbits with positive integers $(M,N)$.

Changing the summation over periodic orbits to the integration over periodic
orbit length with the above density one concludes that the left hand side of 
Eq.~(\ref{25}) in the limit of small $\epsilon$ equals \cite{7}
\begin{equation}
2\pi \epsilon\frac{A^2}{2\pi^2}\int_0^{\infty}
\frac{e^{-\epsilon l/k}}{16A k l}ldl=\frac{A}{16\pi}.
\end{equation}
But this is exactly equal to $\bar{d}(E)$ as it was predicted by the bootstrap
equation (\ref{22}).

Though in this case the computation of the smooth part is simple the
bootstrap formula (\ref{22}) should be valid for oscillatory terms as well
for which the diagonal approximation cannot be applied. It is the calculation of
oscillatory terms (proportional to $\exp (ikl_p)$) which is the main
subject of this Section.

The left-hand side of (\ref{22}) contains a double sum over pairs of periodic 
orbits. Let us try to find saddle points in this double sum. 

A periodic orbit for a rectangular billiard (and for any integrable models
in 2 dimensions (see Section 8)) is defined by two integers $M$ and $N$
(or integer vector $(M,N)$) and its length equals  
the modulus of vector with components $Ma$ and $Nb$,
$L(M,N)=\sqrt{(Ma)^2+(Nb)^2}$. The length of a periodic orbit defined by integers
$M+\delta M$ and $N+\delta N$ has the following expansion up to the second
order in $\delta M$ and $\delta N$
\begin{eqnarray}
&&L(M+\delta M, N+\delta N)\equiv \sqrt{a^2(M+\delta M)^2+b^2(N+\delta N)^2}
\label{28}\\
&&=L(M,N)+\frac{a^2 M\delta M +b^2N \delta N}{L(M,N)}+
\frac{A^2}{2L^3(M,N)}(N\delta M-M\delta N)^2.
\nonumber
\end{eqnarray}
The double sum in Eq.~(\ref{22}) has the form 
$\sum_{p_1,p_2}\exp ik(l_1-l_2)$. Each periodic orbit in this sum is defined
by two integers and the sum is really the sum over 4 integers. 
Let us assume that saddle point manifolds consist of periodic orbits
with integers $(M_1+\delta M_1,N_1+\delta N_1)$ for the first sum and 
$(M_2+\delta M_2,N_2+\delta N_2)$ for the second one with  unknown
integers $(M_i,N_i)$, $(\delta M_i,\delta N_i)$. The necessary condition 
for the existence of a saddle point in the double sum is the cancellation 
of linear terms in the exponent. From Eq.~(\ref{28}) one concludes that the 
saddle point condition has the form 
\begin{equation}
\frac{a^2 M_1\delta M_1+b^2 N_1\delta N_1}{L(M_1,N_1)}-
\frac{a^2 M_2\delta M_2+b^2 N_2\delta N_2}{L(M_2,N_2)}=0.
\label{29}
\end{equation}
The important restriction is that we are looking at the {\bf integer} 
solutions of this equation. We shall also assume that the ratio $a^2/b^2$
is a `good' irrational number (which is a necessary condition that the
spectral statistics  of such billiard will be close to the Poisson statistics
\cite{11}). In this case the 
only possibility to find non-trivial solutions of the above saddle point 
equation is to require  that the lengths $L(M_1,N_1)$ and $L(M_2,N_2)$ are
commensurable
\begin{equation}
\frac{L(M_1,N_1)}{L(M_2,N_2)}=\frac{r_1}{r_2},
\end{equation}
with certain integers $r_1$ and $r_2$. This condition means that saddle
point values of integers $M_i$ and $N_i$ are the following
\begin{equation}
M_1=r_1 m,\; M_2=r_2 m,\; N_1=r_1 n,\; N_2=r_2 n,
\end{equation}
where the pair of integers $m$ and $n$ has no common factors (i.e. they are 
co-prime integers).

Now the saddle point condition (\ref{29}) takes the form
\begin{equation}
a^2 m\delta M_1+b^2n\delta N_1-a^2m\delta M_2-b^2n\delta N_2=0.
\end{equation}
As we have assumed that the ratio $a^2/b^2$ is an irrational number the 
only way to fulfill this equation is to cancel terms in front of $a^2$ and 
$b^2$ separatively
\begin{equation}
\delta M_1-\delta M_2=0,\;\delta N_1-\delta N_2=0.
\end{equation}
This argumentation shows that in the double sum (\ref{22}) there exists
saddle point manifold consisted from  periodic orbits defined by the following 
pairs of integers
\begin{equation}
M_1=r_1m+\delta M,\;N_1=r_1n+\delta N,\;
M_2=r_2m+\delta M,\;N_2=r_2n+\delta N,
\label{35}
\end{equation}
for arbitrary co-prime integers $m$ and $n$ and arbitrary $\delta M$
and $\delta N$.

Note that the knowledge of pairs $(M_1,N_1)$ and $(M_2,N_2)$ uniquely 
defines the pair $(m,n)$ and the difference $r_1-r_2$ provided that it is
non-zero. Namely, $r_1-r_2$ is the greatest common factor of the pair
$(M_1-M_2,N_1-N_2)$ and $(m,n)$ is the ratio of the division of 
$(M_1-M_2,N_1-N_2)$ on $r_1-r_2$.

To find the value of the double sum (\ref{22}) in the saddle point 
approximation it is necessary to perform the summation over all saddle
point manifold defined in Eq.~(\ref{35}) in the Gaussian approximation 
(i.e. expanding all actions up to the quadratic terms in $\delta M$ and 
$\delta N$ as in Eq.~(\ref{28})). 

Let us denote the left-hand part of Eq.~(\ref{22}) by $D(E)$. Using 
(\ref{28}) and (\ref{35}) one obtains
\begin{eqnarray}
&&D(E)=\frac{A^2\epsilon}{8\pi^2 k}\sum_{l_0,r_1,r_2}\sum_{\delta M,\delta N}
\frac{1}{l_0\sqrt{r_1r_2}}\nonumber\\
&&
\exp (ik(r_1-r_2)l_0-(r_1+r_2)\frac{\epsilon l_o}{2k}
+ik\frac{A^2}{2l_0^3}(\frac{1}{r_1}-\frac{1}{r_2})t^2) +c.c.,
\end{eqnarray}
where $t=n\delta M-m\delta N$, and $l_0=\sqrt{a^2m^2+b^2n^2}$.

Note that the quadratic form in the exponent is not positive definite. But
it is easy to check that if the quantity $t$ has two identical values
$$n\delta M_1-m\delta N_1=n\delta M_2-m\delta N_2,$$ 
and integers $m$ and $n$ have no common factor (as it was assumed above)
then $\delta M_2=\delta M_1+lm$ and $\delta N_2=\delta N_1+ln$ for a certain
integer $l$. But from Eq.~(\ref{35}) it follows that this analog of zero modes
corresponds to a change of the repetition number: $r_2=r_2+l$. 
Therefore the restriction of the summation to orbits with fixed repetition
numbers is equivalent to the summation over all values of $t$ only once.

It means
\begin{eqnarray}
D(E)&=&\frac{A^2\epsilon}{8\pi^2 k}\sum_{l_0}\sum_{r_1 r_2=1}^{\infty}
\frac{1}{l_0\sqrt{r_1,r_2}}
\exp (ik(r_1-r_2)l_0-(r_1+r_2)\frac{\epsilon l_0}{2k})
\nonumber\\
&\times &\sum_{t=-\infty}^{\infty}\exp(ik\frac{A^2(r_1-r_2)}{2l_0^3r_1r_2}t^2) +c.c.
\end{eqnarray}
Let $r_1=r_2+r$. Changing the summation over $t$ to the integration one
finds
\begin{equation}
D(E)=\frac{A\epsilon}{8\pi^2 k\sqrt{k}}\sum_{l_0}\sum_{r=1}^{\infty}
\frac{\sqrt{2\pi l_0^3}}{l_0\sqrt{r}}
\exp (ikrl_0-r\frac{\epsilon l_0}{2k}-i\frac{\pi}{4})
\sum_{r_1=0}^{\infty}e^{-r_1 \epsilon l_0/k} +c.c.
\end{equation}
At small $\epsilon$ the last sum tends to $k/(\epsilon l_0)$ and
\begin{equation}
D(E)=\frac{A}{2\pi}\sum_{l_0}\sum_{r=1}^{\infty}
\frac{1}{\sqrt{8\pi k l_0 r}}
\exp (ikrl_0-r\frac{\epsilon}{2k}-i\frac{\pi}{4}) +c.c.
\label{38}
\end{equation}
which is exactly the right hand side of Eq.~(\ref{22}). 

Note that the changing the summation over $t$ to the integration in the sum
$\sum_{t=-\infty}^{\infty}\exp (i\pi x t^2)$ is equivalent to ignoring terms
of the form $e^{-i\pi m^2/x}/\sqrt{x}$ with integer $m\ne 0$. This is 
correct when $x\rightarrow 0$ and only in a weak sence (i.e. after the
multiplication of both sides of the equality
\begin{equation}
\sum_{t=-\infty}^{\infty}e^{i\pi x t^2}-\frac{1}{\sqrt{x}}=
\frac{1}{\sqrt{x}}\sum_{m=-\infty,\;m\ne 0}^{\infty}  e^{-i\pi m^2/x}
\end{equation}
by a suitable chosen test function). It is in such weak sence that one
should treat equalities similar to Eq.~(\ref{38}). This is not a restriction
of our method as all semi-classical trace formulas have mathematical  meaning 
only in a weak sence.

The above calculations clearly demonstrate that the proposed saddle 
point method is sufficient to obtain  the bootstrap condition (\ref{22})
for $n=2$. Below we show that exactly the same considerations permit to 
verify the bootstrap conditions (\ref{23}) for all $n$.

Because $d_{\epsilon}(E)=d_{+\epsilon}(E)+d_{-\epsilon}(E)$,
$d_{\epsilon}^n(E)$ is given by the sum
\begin{equation}
d_{\epsilon}^n(E)=\sum_{\nu_1,\nu_2}C_n^{\nu_1}
d_{+\epsilon}^{\nu_1}(E)d_{-\epsilon}^{\nu_2}(E),
\end{equation}
where $\nu_i\ge 0$ and $\nu_1+\nu_2=n$. Substituting instead of 
$d_{\pm \epsilon}(E)$ its semi-classical expression (\ref{17}) one gets
\begin{equation}
d_{+\epsilon}^{\nu_1}(E)d_{-\epsilon}^{\nu_2}(E)=
    (\frac{A}{2\pi \sqrt{8\pi k}})^{\nu_1+\nu_2}  
\sum_{l,l'}
\frac{e^{ik(L-L')-\epsilon(L+L')/2k-i\pi(\nu_1-\nu_2)/4}}
{\sqrt{(l_1\cdots l_{\nu_1})(l_1' \cdots l_{\nu_2}')}},
\label{43}
\end{equation}
and $L=l_1+\ldots+l_{\nu_1}$, $L'=l_1'+\ldots+l_{\nu_2}'$. The summation here
is performed over all periodic orbits with lengths  $l_j$,  
$j=1,\ldots,\nu_1$ corresponding to terms with positive exponent 
and over orbits with lengths $l_k'$, $k=1,\ldots,\nu_2$ from terms 
with negative exponent. 

Let each  `positive' periodic orbit is defined by a pair of  integers $(M_j,N_j)$ 
and respectively the  `negative' orbit is associated with a pair  $(M_k',N_k')$.
The same arguments as above prove that the saddle point manifold
(i.e. the set of integer vectors for which linear terms in the difference $L-L'$
cancel) has the following form
\begin{equation}
M_j=r_jm+\delta M_j,\;N_j=r_jn+\delta N_j,\;M_k'=r_k'm+\delta M_k',\;
N_k'=r_k'n+\delta N_k',
\label{44}
\end{equation}
with 2 restrictions
\begin{equation}
\sum_{j=1}^{\nu_1}\delta M_j-\sum_{k=1}^{\nu_2}\delta M_k'=0,\;
\sum_{j=1}^{\nu_1}\delta N_j-\sum_{k=1}^{\nu_2}\delta N_k'=0.
\label{45}
\end{equation}
It is convenient to rewrite these conditions in terms of integer vectors 
\begin{equation}
\vec{N}_j=(M_j,N_j),
\end{equation}
associated with each periodic trajectory. Now Eqs.~(\ref{44})-(\ref{45})
take the form
\begin{equation}
\sum_{j=1}^{\nu_1}\vec{N}_j-\sum_{k=1}^{\nu_2}\vec{N}_k'=r\vec{n},
\label{46}
\end{equation}
where 
\begin{equation}
r=\sum_{j=1}^{\nu_1}r_j-\sum_{k=1}^{\nu_2}r_k',
\end{equation}
and
\begin{equation}
\vec{n}=(m,n).
\end{equation}
Because we assume that $m$ and $n$ are co-prime integers they are defined
uniquely from the knowledge of $\vec{N}_j$ and $\vec{N}_k'$ and each term in
the multiple sum in Eq.~(\ref{43}) can be uniquely attributed to the new
summation of the form of Eqs.~(\ref{44})-(\ref{45}).

The total contribution of these saddle points to the $n$-fold sum in 
Eq.~(\ref{43}) in the Gaussian approximation is
\begin{eqnarray}
&&d_{+\epsilon}^{\nu_1}(E)d_{-\epsilon}^{\nu_2}(E)=\sum_{l_0}P^{\nu_1+\nu_2} 
\sum_{R,R'}
\frac{1}{\sqrt{(r_1\cdots r_{\nu_1})(r_1' \cdots r_{\nu_2}')}}\times
\nonumber\\
&&\exp (ik(R-R')l_0-\frac{\epsilon(R+R')l_0}{2k}-i\frac{\pi}{4}(\nu_1-\nu_2))S,
\end{eqnarray}
where
\begin{equation}
P=\frac{A}{2\pi \sqrt{8\pi kl_0}},
\end{equation}
and
\begin{equation}
S=\sum_{ t,t'}\exp (iQ(\frac{t_1^2}{r_1}+\ldots+
\frac{t_{\nu_1}^2}{r_{\nu_1}}-\frac{t_1'^2}{r_1'}-\ldots-
\frac{t_{\nu_2}'^2}{r_{\nu_2}'}))\delta (t_1+\ldots +t_{\nu_1}-t_1'-\ldots -
t_{\nu_2}').
\end{equation} 
Here
\begin{equation}
Q=k\frac{A^2}{2l_0^3},
\end{equation}
$R=r_1+\ldots+r_{\nu_1}$, $ R'=r_1'+\ldots+r_{\nu_2}'$,
$t_j=n\delta M_j-m\delta N_j$, $ t_k'=n\delta M_k'-m\delta N_k'$,
and the $\delta$-function is a consequence of the saddle point conditions
(\ref{45}) in variables $t$ and $t'$.

Changing the summation over $t$ and $t'$ into the integration and using the
usual representation of the $\delta$-function
\begin{equation}
\delta (x)=\frac{1}{2\pi}\int d\alpha e^{i\alpha x},
\end{equation}
one can easily perform all integrals and one obtains
\begin{equation}
S=e^{i\frac{\pi}{4}(\nu_1-\nu_2-\mbox{\small sgn} (r))}
\sqrt{\frac{(r_1\cdots r_{\nu_1})(r_1'\cdots r_{\nu_2}')}{|r|}}
(\frac{\pi}{Q})^{\frac{\nu_1+\nu_2-1}{2}},
\end{equation}
where $r=R-R'$ (note that for oscillating terms $r\ne 0$).

This results leads to the following expression for $d_{\epsilon}^n(E)$
\begin{equation}
d_{\epsilon}^n(E)=\sum_{l_0,r}
P^n(\sqrt{\frac{\pi}{Q}})^{n-1}\sum_{R_1-R_2=r}N(R_1,R_2)
\frac{e^{ikl_0r-i\pi \mbox{\small sgn}(r)/4-\epsilon l_0R_2/k}}{\sqrt{|r|}},
\end{equation}
where
\begin{equation}
N(R_1,R_2)=\sum_{\nu_1=1}^n C_n^{\nu_1}n(R_1,R_2 ;\nu_1),
\end{equation}
and $ n(R_1,R_2 ;\nu_1)$ is the number of terms with fixed number of 
positive, $R_1$, and negative, $R_2$, repetitions.

To find this number let us first compute the number of representation of a given
number $M$ into a sum of $k$ integers: $M=n_1+\ldots+n_k$ and all
$n_i\ge 1$. It is easy to see that this number is the coefficient of $x^M$ in
the expansion $x^k/(1-x)^k$, therefore there exists $C_{M-1}^{k-1}$ ways
of representing $M$ as a sum of $k$ non-zero integers and 
\begin{equation}
N(R_1,R_2)= \sum_{\nu_1=1}^{n-1} 
C_n^{\nu_1}C_{R_1-1}^{\nu_1-1}C_{R_2-1}^{n-\nu_1-1}.
\end{equation}
Because we are interested in the region $R_1, R_2 \rightarrow \infty$ with
fixed difference $R_1-R_2=r$ we can use the following asymptotics of the 
binomial coefficients
$$ C_{R_1-1}^{\nu_1-1}\rightarrow \frac{R_1^{\nu_1-1}}{(\nu_1-1) !}.$$
Finally
\begin{equation}
N(R_1,R_2)= \sum_{\nu_1=1}^{n-1} 
C_n^{\nu_1}C_{n-2}^{\nu_1-1}\frac{R_2^{n-2}}{(n-2) !}.
\end{equation}
The remaining sum
$$\sum_{\nu_1=1}^{n-1} C_n^{\nu_1}C_{n-2}^{\nu_1-1}=
\sum_{\nu_1=1}^{n-1} C_n^{\nu_1}C_{n-2}^{n-\nu_1-1}$$
is the coefficient of $x^{n-1}$ in the expansion $(1+x)^{n}(1+x)^{n-2}$
therefore
\begin{equation}
\sum_{\nu_1=1}^{n-1} C_n^{\nu_1}C_{n-2}^{\nu_1-1}=C_{2n-2}^{n-1}.
\label{51}
\end{equation}
The total contribution is
\begin{equation}
d_{\epsilon}^n(E)=C_{2n-2}^{n-1}\sum_{l_0,r}P^n(\frac{\pi}{Q})^{(n-1)/2}
\frac{e^{ikl_0r-i\pi \mbox{\small sgn}(r)/4}}{\sqrt{|r|}}\sum_{R_2=1}^{\infty}
\frac{R_2^{n-2}}{(n-2)!}e^{-\epsilon l_0R_2/k}.
\end{equation}
When $\epsilon \rightarrow 0$ the last sum tends to $(k/\epsilon l_0)^{n-1}$
and because
\begin{equation}
P\sqrt{\frac{\pi}{Q}}\frac{2k}{l_0}=\frac{1}{2\pi},
\label{53}
\end{equation}
one obtains the following relation 
\begin{equation}
\lim_{\epsilon \rightarrow 0}\frac{(4\pi \epsilon)^{n-1}}{C_{2n-2}^{n-1}}
d_{\epsilon}^n(E)= \frac{A}{2\pi}\sum_{l_p}
\frac{e^{ikl_p-i\pi/4}}{\sqrt{8\pi k l_p}}+c.c.=d^{osc}(E).
\end{equation}
As
\begin{equation}
\frac{4^{n-1}}{C_{2n-2}^{n-1}}=\frac{(2n-2) ! !}{(2n-3) ! !}
\end{equation}
this results coincides with oscillating part of the general bootstrap condition 
(\ref{23}) for all $n$.

These calculations demonstrate that our saddle point method  
reproduces correctly the oscillating part of bootstrap condition but it
remains the question how to apply this  method for the smooth part of the
bootstrap condition with $n>2$.

\section{Saddle points for smooth terms}\label{smooth}

Immediate difficulty of  generalization of the above method to the computation
of  the smooth part of the bootstrap condition (\ref{23}) in the fact that now one 
cannot attribute uniquely values of $(m,n)$ to each term in the multiple sum 
over periodic orbits (\ref{43}). Even for $n=2$ the calculation of the smooth 
term is done by the diagonal approximation and requires  the knowledge of the 
number of  periodic orbits in contrast to the saddle point method discussed
in the above Section.

Each periodic orbit of rectangular billiard is defined by
an unique  vector\footnote{For the purpose of this Section it will be 
  natural to consider vectors with component $(Ma,Nb)$ but to be
  consistent with more general case discussed in Section \ref{general} this
  definion is more convenient.} 
\begin{equation}
\vec{N}=(M,N),
\label{54}
\end{equation}
with positive integers $M$,  $N$ and the length of this periodic orbit is given
by the following expression
\begin{equation}
L(\vec{N})=\sqrt{(aM)^2+(bN)^2}. 
\end{equation}
In the polar coordinates with $0\le \phi \le \pi/2$
\begin{equation}
M=R\cos \phi,\;N=R\sin \phi,
\end{equation}
the local density of periodic orbits $\rho(l,\phi)$ may be computed from the relation
\begin{equation}
\int \rho(l,\phi)dl d\phi\approx 
\int_0^{\infty}dMdN\delta (L(\vec{N})-l)dl,
\end{equation}
which leads to
\begin{equation}
\rho(l,\phi)=\frac{l}{L_0^2(\phi)},
\label{57}
\end{equation}
and 
\begin{equation}
L_0(\phi)=\sqrt{a^2\cos^2 \phi+b^2\sin^2 \phi}.
\label{58}
\end{equation}
As
\begin{equation}
\int_0^{\pi/2}\frac{d\phi}{L_0^2(\phi)}=\frac{\pi}{2ab},
\label{59}
\end{equation}
the integrated density of periodic orbit length
\begin{equation}
\rho(l)=\int_0^{\pi/2}\rho(l,\phi)d\phi =\frac{\pi l}{2ab},
\end{equation}
in agreement with Eq.~(\ref{26}).

To apply the saddle point method it is necessary  to construct a configurations of
vectors (\ref{54}) (saddle point manifold) such that any variations of them will
decrease (or increase) the total length (i.e. the expansion of the total length will
not contain terms linear on variations). From geometrical considerations it is clear 
that the only possibility for such a configuration is the case when all vectors are 
parallel. To build the saddle point manifold we shall proceed as follows.

Let us consider $n$ vectors $\vec{N}_i$ $i=1,\ldots,n$ in polar coordinates 
\begin{equation}
\vec{N}_i=R_i(\cos \phi_i,\sin \phi_i).
\end{equation}
The condition that all these vectors are almost parallel is equivalent to
the statement that all polar angles are close to each other
\begin{equation}
\phi_i=\phi+\delta \phi_i,
\end{equation}
and $\delta \phi_i\ll \phi$. Under these conditions $L(\vec{N}_i)$ can be
calculated from Eq.~(\ref{28}) up to the second order of $\delta \phi_i$
\begin{equation}
L(\vec{N}_i)=R_iL_0+\frac{(b^2-a^2)\sin 2\phi}{2L_0}R_i\delta \phi_i 
+\frac{A^2}{2L_0^3}R_i(\delta \phi_i)^2.
\label{60}
\end{equation}
Let us now compute the difference of actions
\begin{equation}
S_n=k(\sum_{i=1}^nL(\vec{N}_i)-L(\sum_{i=1}^n\vec{N}_i)).
\end{equation}
Note that this operation is possible because the sum of integer vectors
(\ref{54}) with positive components is also an integer vector with positive
components and therefore it defines a periodic orbit. Direct application of
Eq.~(\ref{60}) gives
\begin{eqnarray}
S_n&=&Q(\sum_{i=1}^nR_i(\delta
\phi_i)^2-\frac{1}{R_1+\ldots+R_n}(\sum_{i=1}^nR_i\delta \phi_i)^2)
\nonumber \\
&=&\frac{Q}{R_1+\ldots+R_n}\sum_{i<j}R_iR_j(\phi_i-\phi_j)^2,
\label{64}
\end{eqnarray}
where
\begin{equation}
Q=\frac{k A^2}{2L_0^3(\phi)}.
\end{equation}
As in the previous Section
$$ d_{\epsilon}^n(E)=\sum_{\nu_1+\nu_2=n}C_n^{\nu_1} 
d_{+\epsilon}^{\nu_1}(E) d_{-\epsilon}^{\nu_2}(E),$$
and under the same notation as above
\begin{eqnarray}
d_{+\epsilon}^{\nu_1}(E) d_{-\epsilon}^{\nu_2}(E)&=&
(\frac{A}{2\pi \sqrt{8\pi k}})^{\nu_1+\nu_2}\sum_{l,l'}
\frac{1}{\sqrt{(l_1\cdots l_{\nu_1})(l_1' \cdots l_{\nu_2}')}}\times 
\nonumber\\
&&\exp (ik(L-L')-\frac{\epsilon(L+L')}{2k}-i\frac{\pi}{4}(\nu_1-\nu_2)),
\label{61}
\end{eqnarray}
with $L=l_1+\ldots+l_{\nu_1}$ and  $L'=l_1'+\ldots+l_{\nu_2}'$.

Each periodic orbit is defined by an integer vector $\vec{N}_i$ and 
according to the above discussion the saddle point manifold for the smooth
part of this expression is constructed from vectors $\vec{N}_i$ and 
$\vec{N}_j'$ which are almost parallel each to other and obey the following
relation (because it should be the smooth part)
\begin{equation}
\sum_{j=1}^{\nu_1}\vec{N}_j=\sum_{k=1}^{\nu_2}\vec{N}_k'.
\label{62}
\end{equation}
One may easily check that the regrouping terms in Eq.~(\ref{61}) according to
the sum of vectors $\vec{N}_j$ and that of  $\vec{N}_k'$ avoids the double
counting and permits the correct arrangement of different contributions.

In polar coordinates the saddle point condition (\ref{62})  means that 
\begin{equation}
R_1+\ldots+R_{\nu_1}=R_1'+\ldots+R_{\nu_2}',
\end{equation}
and 
\begin{equation}
R_1\delta \phi_1+\ldots+R_{\nu_1}\delta \phi_{\nu_1}=
R_1'\delta \phi_1'+\ldots+R_{\nu_2}'\delta \phi_{\nu_2}'.
\end{equation}
From general considerations (and  from Eq.~(\ref{64})) it  follows that the
difference of lengths $(L-L')$ depends only on differences of angles between
different vectors and therefore one of these angles (say $\phi_1=\phi$) can
be chosen arbitrary and under the condition of smallness of these differences
the sum over periodic orbits in Eq. (\ref{61}) can be substitute by the integration
over components of $\vec{N}_i$  and Eq. (\ref{61}) can be transformed as follows
\begin{eqnarray}
&&d_{+\epsilon}^{\nu_1}(E) d_{-\epsilon}^{\nu_2}(E)=
\int_0^{\pi/2} d\phi P^{(\nu_1+\nu_2)}
\int \prod_{j=1}^{\nu_1}dR_j\prod_{k=1}^{\nu_2}dR_k'
\prod_{j=2}^{\nu_1} d\phi_j\prod_{k=1}^{\nu_2} d\phi_k'\times
\nonumber \\
&&\delta(R-R')\delta (R\phi-R' \phi')
\sqrt{(R_1\cdots R_{\nu_1})(R_1' \cdots R_{\nu_2}')}\times\\
&&\exp (iQ(\sum_{j=2}^{\nu_1}R_j(\phi_j)^2 -
\sum_{k=1}^{\nu_2}R_k'(\phi_k')^2)
-\frac{(R+R')L_0}{2k}-i\frac{\pi}{4}(\nu_1-\nu_2)),
\nonumber
\end{eqnarray}
where $R=R_1+\ldots+R_{\nu_1}$,  $R'=R_1'+\ldots+R_{\nu_2}'$,
$R \phi=R_2\phi_2+\ldots+R_{\nu_1}\phi_{\nu_1}$,
$R'\phi'=R_1'\phi_1'+\ldots+R_{\nu_2}'\phi_{\nu_2}'$, and 
\begin{equation}
P=\frac{A}{2\pi \sqrt{8\pi kL_0(\phi)}}.
\end{equation}
Note the absence of $\phi_1$ in these expressions.

Using the standard representation of the $\delta$-function
\begin{equation}
\delta (R \phi-R'\phi')=\frac{1}{2\pi}\int_{-\infty}^{\infty}
e^{i\alpha(R \phi-R' \phi')},
\label{72}
\end{equation}
one can easily perform the angular integration over all independent 
angles $\phi_j$ and $\phi_k'$
\begin{equation}
<\sum_{l,l'}e^{i(L-L')}\delta (R \phi-R' \phi')>=
(\frac{\pi}{Q})^{(\nu_1+\nu_2)/2-1}\frac{e^{i\pi(\nu_1-\nu_2)/4}}
  {\sqrt{(R_1\cdots R_{\nu_1})(R_1'\cdots R_{\nu_2}')}}.
\end{equation}
Combining this result together with Eq.~(\ref{61}) one obtains
\begin{eqnarray}
<d_{+\epsilon}^{\nu_1}(E)d_{-\epsilon}^{\nu_2}(E)>=&&
\int_0^{\pi/2} d\phi P^{(\nu_1+\nu_2)}(\frac{\pi}{Q})^{(\nu_1+\nu_2)/2-1}
\times \nonumber\\
&&\int_0^{\infty} \prod_{j=1}^{\nu_1}dR_j\prod_{k=1}^{\nu_2}dR_k'
\delta(R-R')e^{-(R+R')L_0/2k}.
\end{eqnarray}
The  integrals over $R$ and $R'$ may be computed either by using the
representation (\ref{72}) of the $\delta$-function  or by the 
integration first by one variable e.g. $R_{\nu_2}'$.
Performing the change of variables $R_j=k\tau_j/L_0$,  $R_k'=k\tau_k'/L_0$ and
taking into account that $R_{\nu_2}'\ge 0$ one obtains
\begin{eqnarray}
&&\int_0^{\infty} \prod_{j=1}^{\nu_1}dR_j\prod_{k=1}^{\nu_2}dR_k'
\delta(R-R')e^{-(R+R')L_0/2k}
\nonumber\\
&&=(\frac{k}{L_0})^{(\nu_1+\nu_2-1)}
\int_0^{\infty} {\cal D}\tau {\cal D} \tau'\theta(\tau-\tau')e^{-\epsilon \tau},
\end{eqnarray}
where $\tau=\tau_1+\ldots+\tau_{\nu_1}$,
$\tau'=\tau_1'+\ldots+\tau_{\nu_2-1}'$, 
${\cal D}\tau=d\tau_1\cdots\tau_{\nu_1}$, and 
${\cal D}\tau'=d\tau_1'\cdots\tau_{\nu_2-1}'$.

The integral over ${\cal D}\tau'$ has the form
\begin{equation}
g(\tau)=\int_0^{\infty}d\tau_1'\cdots\tau_{\nu_2-1}'
  \theta(\tau-\tau_1'+\ldots+\tau_{\nu_2-1}'),
\end{equation}
and may be computed by the Laplace transform over variable $\tau$. The
result is
\begin{equation}
g(\tau)=\frac{\tau^{\nu_2-1}}{(\nu_2-1)!}.
\end{equation}
The same method  applied to the integral over ${\cal D}\tau$ gives 
\begin{eqnarray}
&&\int {\cal D}\tau
     {\cal D} \tau'\theta(\tau-\tau')e^{-\epsilon \tau}=
\int d\tau g(\tau) e^{-\epsilon \tau}\int d\tau_1\cdots\tau_{\nu_1} 
\delta(\tau-\tau_1-\ldots-\tau_{\nu_1})\nonumber\\
&&=
\int_0^{\infty} d\tau \frac{\tau^{\nu_1+\nu_2-2}}{(\nu_1-1)!(\nu_2-1)!}
e^{-\epsilon \tau}=
\frac{(\nu_1+\nu_2-2)!}{(\nu_1-1)!(\nu_2-1)!}\epsilon^{-(\nu_1+\nu_2-1)}.
\end{eqnarray}
Taking into account Eq.~(\ref{59}) and that (compare with Eq.~(\ref{53}))
\begin{equation}
P\sqrt{\frac{\pi}{Q}}\frac{2k}{L_0}=\frac{1}{2\pi},
\label{79}
\end{equation}
and
\begin{equation}
\int_0^{\pi/2}P\sqrt{\frac{Q}{\pi}}d\phi=\frac{A}{16\pi}=\bar{d},
\label{79b}
\end{equation}
one obtains
\begin{equation}
<d_{+\epsilon}^{\nu_1}(E)d_{-\epsilon}^{\nu_2}(E)>=  
C_{\nu_1+\nu_2-2}^{\nu_1-1}
\frac{A}{(4 \pi \epsilon)^{n-1}16\pi},
\label{80}
\end{equation}
and
\begin{equation}
<d_{\epsilon}^n>=
\frac{A}{(4 \pi \epsilon)^{n-1}16\pi}\sum_{\nu_1=1}^{n-1}C_n^{\nu_1}C_{n-2}^{\nu_1-1}.
\end{equation}
But this sum equals $C_{2n-2}^{n-1}$ (see Eq.~(\ref{51})) and 
\begin{equation}
\lim_{\epsilon \rightarrow 0}\frac{(4\pi \epsilon)^{n-1}}{C_{2n-2}^{n-1}}
<d_{\epsilon}^n(E)>= \frac{A}{16\pi}=\bar{d},
\end{equation}
which agrees with the mean part of the bootstrap condition (\ref{23}).

\section{Mean value of Green function products}

Problems similar to the ones described above appear also in the computation of
mean value of products of the advanced and retarded Green functions of the type
\begin{equation}
<G_{+}^m(E_1)G_{-}^n(E_2)>,
\label{82}
\end{equation}
where $G_{\pm}(E)=G(E\pm i\epsilon)$ and $G(E)$ is the Green function of 
an integrable model. A typical example is  the semi-classical
computation of spectral correlation functions in the Seba billiard
\cite{12,13} or, more generally, in integrable models with diffraction centers.

For clarity  as in the previous Sections let us consider  a rectangular billiard
with periodic boundary conditions. The Green function for this problem has the 
form
\begin{equation}
  G_{\pm}(\vec{x},E)=\frac{1}{A}\sum_n\frac{e^{i\vec{k}_n\vec{x}}}
{E-e_n\pm i\epsilon},
\label{83}
\end{equation}
where  $e_n=k_n^2$ are eigenvalues of the billiard problem,
$\vec{k}_n$ is the vector with components $2\pi (m/a, n/b)$ with integers
$m$, $n$, and $A$ is the area of the rectangle. This Green function permits
also an exact representation as a sum over all classical orbits
connecting points $0$ and $\vec{x}=(x,y)$
\begin{equation}
G_{+}(\vec{x},E)=-\frac{i}{4}\sum_{m,n=-\infty}^{\infty}
H_0^{(1)}(k\sqrt{(x+ma)^2+(y+nb)^2}),
\label{84}
\end{equation}
and $H_0^{(1)}(x)$ is the Hankel function,
$G_{-}(\vec{x},E)=G_{+}^*(\vec{x},E)$.

In diffraction problems one is interested in the Green function
when $x\rightarrow 0$. In this limit the Green function diverges and
requires a cut-off at small $x$ (i.e. a regularization).
The divergent part comes  only from the contribution of the shortest trajectory
with $m=n=0$. Taking into account the asymptotic behavior of the Hankel
function at small $x$ one concludes that the `renormalized' Green function
(\ref{84}) (when $|x|\rightarrow \mu$ and $\mu$ is small ) is a sum of 2 terms
\begin{equation}
G_{+}(E)=\bar{g}_{+}(E) +g_{+}^{osc}(E),
\label{84b}
\end{equation}
where (in 2 dimensions)
\begin{equation}
\bar{g}_{+}(E)=\frac{1}{2\pi}\ln (k\mu) - \frac{i}{4} 
\label{84c}
\end{equation}
and
\begin{equation}
g_{+}^{osc}(E)=-\frac{i}{4}
\sum_{p} H_0^{(1)}(kL_{p}),
\label{85}
\end{equation}
where $L_p$ denotes the length of a periodic orbit defined by 2 integers
$m$ and $n$, $L_{p}=\sqrt{(ma)^2+(nb)^2}$, the summation is done over all
$m,n$ except $m=n=0$, and  parameter $\mu$ is a renormalization parameter (a
cut-off at small distances). 
In the quantization of the Seba billiard
the necessity of the renormalization is connected with the fact that the
$\delta$-function potential is too singular in dimensions $\ge 2$ and
Eq.~(\ref{85}) defines one parameter self-adjoint extension (defined by
$\mu$) of a singular Hamiltonian \cite{12}. 
Note that $<G_{\pm}>=\bar{g}_{\pm}(E)$ and $<g^{osc}_{\pm}(E)>=0$.

The semi-classical approximation corresponds to $k\rightarrow \infty$ and
the oscillating part of the Green function takes the form
\begin{equation}
g_{+}^{osc}(E)=\sum_{p}
   \frac{e^{ikL_{p}-3\pi i/4-\epsilon L_{p}/2k}}{\sqrt{8\pi k L_p}} .
\label{86}
\end{equation}
Let 
\begin{equation}
g_{mn}(E_1,E_2)=\frac{1}{A^{m+n}}\sum_n 
\frac{1}{(E_1-e_n+i\epsilon)^m(E_2-e_n-i\epsilon)^n}.
\end{equation}
Assuming that all energy levels have the Poisson distribution (i.e. they are
independent random variables with mean density $\bar{d}$) it follows that
\begin{eqnarray}
g_{mn}(E_1,E_2)&=&\frac{1}{A^{m+n}}\int 
\frac{\bar{d}de }{(E_1-e+i\epsilon)^m(E_2-e-i\epsilon)^n}
\nonumber\\
&=&\frac{2\pi i \bar{d}(-1)^{n-1}}{A((E_1-E_2) A)^{m+n-1}}
C_{n+m-2}^{n-1}.
\end{eqnarray}
For rectangular billiard $\bar{d}=A/(16\pi)$ and
\begin{equation}
g_{mn}(E_1,E_2)=
\frac{i(-1)^{n-1}}{8((E_1-E_2) A)^{m+n-1}}C_{n+m-2}^{n-1}.
\label{87}
\end{equation}
It is easy to check (see below) that the most divergent part of
the product (\ref{82}) when $E_2\rightarrow E_1$ equals $g_{mn}(E_1,E_2)$
\begin{equation}
<G_{+}^m(E_1)G_{-}^n(E_2)>\rightarrow g_{mn}(E_1,E_2), 
\mbox{ when }\;\; E_2\rightarrow E_1.
\label{103}
\end{equation}
The purpose of this section is to check that exactly the same answer can be
obtained by using the semi-classical expression (\ref{86}) for  the Green
function and applying the saddle point method discussed in the previous
Sections. The differences between  the computation of
$<d_{+\epsilon}(E)^{\nu_1}d_{-\epsilon}(E)^{\nu_2}>$ performed in the last
Section and that of $<G_{+}^m(E_1)G_{-}^n(E_2)>$ are (i) the absence of
factor $A/2\pi$ and (ii) the change of the definition of $\epsilon$
$\epsilon \rightarrow -i(E_1-E_2)/2$ (and, of course, it is necessary to
substitute $\nu_1\rightarrow m$ and $\nu_2 \rightarrow n$). From Eq.~(\ref{80})
one concludes that the saddle point method gives
\begin{eqnarray}
<G_{+}^m(E_1)G_{-}^n(E_2)>&=&C_{n+m-2}^{n-1}
\frac{2^{m+n-1}A}{16\pi(i(E_2-E_1))^{m+n-1}}(\frac{2\pi}{A})^{m+n}
\nonumber\\
&=&i(-1)^{n-1}C_{n+m-2}^{n-1}
\frac{1}{8((E_1-E_2) A)^{m+n-1}},
\end{eqnarray}  
which agree with the value (\ref{103})  obtained by  direct calculations.

Till now we have considered only the dominant term of the mean value
of Green functions product (\ref{82}) when  $E_2\rightarrow E_1$ (i.e. we
look for contributions with the highest negative power of the difference of energies). 
But this restriction is not necessary and other terms can also be computed by 
the same method. 

The $m$-th power of th e Green function (\ref{83}) may be written as follows
\begin{eqnarray}
G_{+}^m(E)&=&\frac{1}{A^m}\sum_{m_i}\frac{m!}{m_1!m_2!\cdots m_p!}\\
&\times& \sum_{e_j}
\frac{1}{(E-e_1+i\epsilon)^{m_1}(E-e_2+i\epsilon)^{m_2}\cdots
  (E-e_p+i\epsilon)^{m_p}},
\nonumber 
\end{eqnarray}
where the first sum is taken over all positive integers $m_i$ whose sum
equals $m$, $m_1+\ldots +m_p=m$, and the second summation is performed over
different energy eigenvalues $e_j$ such that $e_1\le e_2\le \ldots $. 
Representing $G_{-}^n(E)$ in the similar way and taking into account that
the connected part of the Green function products should
not include mean values of individual Green functions one concludes that
\begin{eqnarray}
&&<G_{+}^m(E_1)G_{-}^n(E_2)>^{c}=
\sum_{p=1}^{\mbox{\small min}(m,n)}\frac{1}{p!} \label{105}\\
&&\times \sum_{m_i,n_i}\left (\frac{m!}{m_1!m_2!\cdots m_p!}\right ) \left (
\frac{n!}{n_1!n_2!\cdots n_p!}\right )
g_{m_1 n_1}g_{m_2 n_2}\cdots g_{m_p n_p},
\nonumber
\end{eqnarray}
where $g_{m n}=g_{mn}(E_1,E_2)$ is given by Eq.~(\ref{87})
and the summation is performed over all positive integers $m_i\ge 1$,
$n_i\ge 1$ such that $\sum_{i=1}^pm_i=m$ and $\sum_{i=1}^p n_i=n$.
The factor $1/p!$ appears from the assuming ordering of variables $e_i$.
Because of the symmetry with  respect to permutations this factor can be
removed by assuming the ordering of (say) $m_i$ variables and the above
expression may be rewritten as follows
\begin{eqnarray}
<G_{+}^m(E_1)G_{-}^n(E_2)>^{c}&=&\sum_{p=1}^{\mbox{\small min}(m,n)}
\sum_{\tilde{m}_i,n_i}\left (\frac{m!}{m_1!m_2!\cdots m_p!}\right ) \left (
\frac{n!}{n_1!n_2!\cdots n_p!}\right )
\nonumber\\
&\times& g_{m_1 n_1}g_{m_2 n_2}\cdots g_{m_p n_p},
\label{105b}
\end{eqnarray}
where $m_1\le m_2 \ldots \le m_p$.

Each $g_{m_i n_i}$ is proportional to
$((E_1-E_2)A)^{1-m_i-n_i}$, therefore  terms with fixed $p$
in  these sums will be proportional to $((E_1-E_2)A)^{p-m-n}$ and
terms with higher values of $p$ correspond to the expansion of
$<G_{+}^m(E_1)G_{-}^n(E_2)>^{c}(E_1-E_2)^{m+n}$ on positive powers 
of $(E_1-E_2)A$.

The above expression represents the value of the connected part of the mean
value of Green function product calculated from the assumption of the
Poisson distribution of energy eigenvalues. The total answer includes also terms
coming from the mean value of the Green function itself (see Eq.~(\ref{84c}))
\begin{equation}
<G_{+}^m(E_1)G_{-}^n(E_2)>=\sum_{k,p}C_m^kC_n^p \bar{g}_{+}^k\bar{g}_{-}^p  
<G_{+}^k(E_1)G_{-}^p(E_2)>^{c},
\end{equation}
and due to Eq~(\ref{105}) this expression can also be organized  as a series on
decreasing powers of $(E_1-E_2)A$.

Let us check that this answer may also be obtained by the saddle point
method. As was demonstrated in Section \ref{smooth} the saddle point
manifold for smooth terms consists of integer vectors $\vec{N}_i$ 
and $\vec{N}_j'$ such that 
\begin{equation}
\sum_{i=1}^m\vec{N}_i=\sum_{j=1}^n\vec{N}_j'.
\label{106}  
\end{equation}
The dominant contribution discussed in the previous Sections corresponds
to the integration over all possible deviations on this manifold. But by
fixing certain deviations it is possible to find lower dimensional manifolds
and the integration over them will give corrections to the dominant result. 

Let as above $n_i\ge 1$ and $m_i\ge 1$ be partitions of $n$ and $m$ 
into positive integers: $n=n_1+\ldots +n_p$ and  $m=m_1+\ldots +m_p$. Each

such partition gives rise to a possible regrouping of integer vectors
\begin{eqnarray}
  \sum_{i=1}^{m_1}\vec{N}_i&=&\sum_{j=1}^{n_1}\vec{N}_j\nonumber\\
  \sum_{i=1}^{m_2}\vec{N}_{m_1+i}&=&\sum_{j=1}^{n_2}\vec{N}_{n_1+j}\nonumber\\
  \ldots & & \ldots \\
  \sum_{i=1}^{m_p}\vec{N}_{m_1+\ldots +m_{p-1}+i}&=&
  \sum_{j=1}^{n_p}\vec{N}_{n_1+\ldots+n_{p-1}+j}.\nonumber
\end{eqnarray}
Such manifold is a part of co-dimension $(p-1)$ of the full saddle point
manifold (\ref{106}) and the summation over all possible deviations on it
will evidently give $g_{m_1 n_1}g_{m_2 n_2}\cdots g_{m_p n_p}$. As there
exists $m!/(m_1!\ldots m_p!)$ possible partitions with fixed $m_i$ and 
$n!/(n_1!\ldots n_p!)$ possible partitions with fixed $n_i$ the total
contribution will be equal exactly to Eq.~(\ref{105b}) which demonstrates
that all correction terms can be computed correctly by the saddle point method. 

Let us note that in problems considered there exist 4 parameters with
dimensions of energy. One is the energy itself or more precisely
the center of energy window which defines  mean values, the second one is
the width of energy window, the third is
the difference of 2 energies, $\epsilon=E_1-E_2$, and the forth is the mean
distance between quantum levels equal $1/\bar{d}$ where $\bar{d}$ is the
mean density of levels. In the end of this Section we discussed the
corrections corresponding to powers of the most important dimensionless parameter
$\bar{d}\epsilon$. All other dimensionalless parameters are assumed to be
large and we consider only the first terms of the expansion on them. The
computation of higher order terms of these parameters, coming e.g. from
corrections to the Gutzwiller trace formulas, are also possible but is
beyond the scope of this paper.

\section{Correlation functions of classical actions}\label{correlation}

It is well known \cite{14}, \cite{15} that there exists a duality between 
quantum spectral statistics and statistics of classical actions. If one
is known the other can be computed by the Fourier transformation.

In this Section we demonstrate that 
the results of the previous Sections can  be rewritten in the form of
correlation functions of classical actions. Let us consider for simplicity the
2-point correlation function of periodic orbit length for rectangular billiard
(calculation of other correlation functions will be discussed elsewhere
\cite{17}) 
\begin{equation}
R_2(s)=\sum_{l,l'}\delta(s-l+l')a(l,l'),
\label{200}
\end{equation}
where $a(l,l')$ is a certain weighted function which should dominates by
large values of $l$ and $l'$. A typical example is 
\begin{equation}
a(l,l')=\delta(L-\frac{l+l'}{2})
\label{113a}
\end{equation}
and $L$ is assumed to be a large quantity. The correlation function with this 
weight will be denoted by $R_2(s,L)$
\begin{equation}
R_2(s,L)=\sum_{l,l'}\delta(s-l+l')\delta(L-\frac{l+l'}{2}).
\end{equation}
It is this function which appears naturally in many applications (see below).

The summation in (\ref{200}) is done over pairs of periodic orbit lengths.
We  implicitly assume that all values of $l$ and $l'$ are permitted.
There are few other possibilities e.g. one can choose  
$l'\ge l$, or  only $l'>l$ which lead to slight modifications of the formulas below.

One has
\begin{equation}
R_2(s)=\frac{1}{2\pi}\int_{-\infty}^{\infty}d\tau f(\tau)e^{i\tau s},
\end{equation}
where
\begin{equation}
f(\tau)=\sum_{l,l'}e^{i\tau(l'-l)}a(l,l').
\end{equation}
This sum splits into 2 parts
\begin{equation}
f(\tau)=f^{(diag)}(\tau)+f^{(osc)}(\tau),
\end{equation}
where
\begin{equation}
f^{(diag)}(\tau)=\sum_{l=l'}a(l,l')\approx \bar{\rho} \int la(l,l)dl,
\end{equation}
where we have used the mean density of periodic orbits (\ref{26}) with
$\bar{\rho}=\pi/2A$ and
\begin{equation}
f^{(osc)}(\tau)=\sum_{l\ne l'}e^{i\tau(l'-l)}a(l,l').
\end{equation}
To compute this function we shall use the saddle point method developed in
the previous Sections. As above the dominant contribution will come from the
saddle point manifold (\ref{35}) and 
\begin{equation}
f^{(osc)}(\tau)=\sum_{l_0}\sum_{r_1,r_2}\int dt 
    \exp (i\tau l_0 (r_2-r_1)+iQ t^2(\frac{1}{r_2}-\frac{1}{r_1}))
    a(r_1l_0, r_2l_0),
\end{equation}
where $Q=A^2\tau /(2l_0^3)$. Performing the integration  and taking into account 
that $r_1,r_2\rightarrow \infty$  but $r_2-r_1=r$ is fixed one obtains
\begin{equation}
f^{(osc)}(\tau)=\sum_{l_0,r}D 
    e^{i\tau l_0 r-i\pi\mbox{\small sgn}(\tau r)/4}\sqrt{\frac{\pi}{Q|r|}},
\end{equation}
where 
\begin{equation}
D=\int dr_1r_1a(r_1l_0, r_1l_0)=\frac{1}{l_0^2}\int ldl a(l,l).  
\end{equation}
Because
\begin{equation}
\sqrt{\frac{\pi}{Q|r|}}\frac{1}{l_0^2}=
\frac{A\bar{\rho}}{2\pi \bar{d}\sqrt{8\pi \tau l_0 r}},
\end{equation}  
we finally get
\begin{equation}
  f(\tau)=\frac{\bar{a}}{\bar{d}}(\bar{d}+\frac{A}{2\pi}\sum_{l_p}
  (\frac{e^{i\tau l_p -i\pi/4}}{\sqrt{8\pi \tau l_p}}+c.c.)),
\label{210}
\end{equation}
where
\begin{equation}
\bar{a}=\sum_l a(l,l)=\bar{\rho} \int la(l,l)dl.
\end{equation}
But the sum in Eq.~(\ref{210}) is exactly the semi-classical expansion of
oscillating part of the trace formula for rectangular billiard (see
Eq.~(\ref{17})) with energy equals $\tau^2$ which means
that 2-point correlation form-factor for periodic orbit lengths is
proportional to the semi-classical trace formula
\begin{equation}
f(\tau)=\frac{\bar{a}}{\bar{d}}d(\tau^2). 
\label{220}
\end{equation}
In particular for rectangular billiard with weighted function (\ref{113a}) 
\begin{equation}
R_2(s,L)=\frac{4\pi}{A^2}L\int d\tau d(\tau^2) e^{i\tau s}
\end{equation}
These results can be considered as a generalization of results of
Refs.~\cite{14}, \cite{15}. In these papers only the smooth part of this
relation has been considered $\bar{f}(\tau)=\bar{d}(\tau^2)$. But  relation 
(\ref{220}) contains more information. In particular it states that the
Fourier transform of classical 2-point correlation function of periodic
orbit lengths for rectangular billiard should have peaks exactly at
eigenvalues of quantum problem with  Newmann boundary conditions. The
generalization of this relation to general integrable models is performed in
the next Section. Numerical calculations for different integrable billiards
confirm well this prediction \cite{16}.

The knowledge of the 2-point correlation function of classical actions
permits to compute easily any quantities depending on a product of 2
densities. For example, 
\begin{eqnarray}
d_{\epsilon}^2(E)&=&2(\frac{A}{2\pi})^2\sum_{l,l'}
\frac{e^{ik(l-l')-\epsilon (l+l')/2k}}{8\pi k \sqrt{ll'}}=
\frac{A^2}{16\pi^2k}\int_{-\infty}^{\infty}ds e^{iks-\epsilon L/k}
R_2(s,L)\frac{dL}{L}
\nonumber\\
&=&\frac{1}{2\pi k}d(E)\int_0^{\infty}e^{-\epsilon L/k}
dL=\frac{1}{2\pi  \epsilon}d(E),
\end{eqnarray}
which coincides with the bootstrap condition (\ref{22}).

It is also instructive to compute the 2-point correlation function of energy
levels
\begin{eqnarray}
R_2(e)&=&<d(E+\frac{e}{2})d(E-\frac{e}{2})>\nonumber\\
&=&<d^2(E)>+(\frac{A}{2\pi})^2
<\sum_{l,l'}\frac{e^{ik(l-l')+
    e (l+l')/4k}}{8\pi k \sqrt{ll'}}+c.c.>.
\end{eqnarray}
The last sum equals 
\begin{equation}
(\frac{A}{2\pi})^2\int_{-\infty}^{\infty}\frac{dL}{8\pi kL}e^{i \frac{e
    L}{2k}}
\int_{-\infty}^{\infty}ds e^{iks}R_2(s,L)
=\frac{1}{4\pi k}d(E)\int_{-\infty}^{\infty}
e^{i\frac{e L}{2k}}dL
=\delta (e) d(E),
\end{equation}
and  the 2-point correlation function for energy levels
of rectangular billiard (and for general integrable systems as well)  has the 
following form
\begin{equation}
R_2(e)=<d^2(E)>+\delta (e) <d(E)>.
\end{equation}
Here the brackets denote a {\bf local} average over a certain energy window.
When this window is very large $<d(E)>=\bar{d}$ but in general this
smoothing defines a new local scale and $<d(E)>$ can deviates from $\bar{d}$.
This result is exactly as was expected. The energy levels of typical integrable
systems (in particular for rectangular billiard) are distributed as independent
random variables but at the {\bf local} scale (or equivalently only after unfolding).
The correlation function of classical lengths (\ref{220}) does the required
transformation from global scale to the local one.

\section{General integrable systems}\label{general}

In the previous Sections for clarity we discussed the case of rectangular
billiard but the method used is not restricted only to this example
and  can be generalized  for general integrable systems as well. 

The starting point of the derivation of the trace formula for integrable
systems is a formal  expression of a Hamiltonian as a function of action
variables \cite{4} $I_i$
\begin{equation}
E=H(I_1,I_2).
\end{equation}
Semi-classical quantization conditions consist in fixing the values of these
action variables \cite{1}-\cite{4}
\begin{equation}
I_i=\hbar (n_i+\frac{1}{4}\mu_i),
\end{equation}
where $n_i$ are integers and $\mu_i$ are certain phases connected with the
type of classical motion (the Maslov indices). Therefore for 2-dimensional
integrable systems (multi-dimensional case will be discussed elsewhere
\cite{17}) the quantum energies are
\begin{equation}
E_{n_1 n_2}=H(n_1+\frac{1}{4}\mu_1,n_2+\frac{1}{4}\mu_2),
\end{equation}
and the quantum density of states is
\begin{equation}
d(E)=\sum_{n_1,n_2}\delta (E-E_{n_1 n_2}).
\end{equation}
Using the Poisson summation formula one obtains
\begin{equation}
d(E)=\sum_{N_1,N_2} \int e^{2\pi i(N_1n_1+N_2n_2)} 
\delta (E-E_{n_1 n_2})dn_1dn_2.
\end{equation}
Separating the term with $N_1=N_2=0$ and setting 
$n_i=I_i/\hbar-\mu_i/4$ leads to
\begin{equation}
d(E)=\bar{d}(E)+d^{osc}(E),
\end{equation}
where 
\begin{equation}
\bar{d}(E)=\frac{1}{\hbar^2}\int \delta (E-H(I_1,I_2))dI_1 dI_2=  
\frac{1}{(2\pi\hbar)^2}\int \delta (E-H(\vec{p},\vec{q}))d\vec{p}d\vec{q},
\end{equation}
and
\begin{equation}
d^{osc}(E)=\frac{1}{\hbar^2}\sum_{N_1,N_2}e^{-i(N_1\mu_1+N_2\mu_2)/2}
\int e^{2\pi i(N_1 I_1+N_2I_2)/\hbar}\delta (E-H(I_1,I_2))dI_1 dI_2.
\end{equation}
It is convenient instead of variables $\vec{I}$ to use variables $e$ and $t$
defined by 
\begin{equation}
\frac{\partial \vec{I}}{\partial e}=\vec{\nu},\;\;
\frac{\partial \vec{I}}{\partial t}=\vec{\tau},
\end{equation}
where the unit vectors $\vec{\nu}$ and $\vec{\tau}$ have the following
components 
\begin{equation}
\nu_i=\frac{1}{\sqrt{\omega_1^2+\omega_2^2}}\omega_i,\;\;
\tau_i=\frac{1}{\sqrt{\omega_1^2+\omega_2^2}}\sum_{j=1}^2e_{ij}\omega_j,
\end{equation}
and
\begin{equation}
\omega_i=\frac{\partial H(I_1,I_2)}{\partial I_i}
\end{equation}
are frequencies of the classical motion,
and $e_{ij}$ is $2\times 2$ antisymmetric tensor with components
$e_{11}=e_{22}=0$ and $e_{21}=-e_{12}=1$.

Vector $\vec{\nu}$ is the unit vector perpendicular to the line of fixed
energy, vector $\vec{\tau}$ is the unit vector tangent to this line,
and $\vec{\nu}^2=\vec{\tau}^2=1$, $\vec{\nu}\vec{\tau}=0$

It is easy to check that $dI_1dI_2=dedt$ and 
\begin{equation}
\int \delta (E-H(e))de=\frac{1}{\sqrt{\omega_1^2+\omega_2^2}}.
\end{equation}
The integral over the surface of constant energy 
\begin{equation}
\int e^{2\pi i(N_1 I_1+N_2I_2)/\hbar}dt  
\end{equation}
can be taken by saddle point method. The saddle points are points where
\begin{equation}
\sum_i N_i\frac{\partial I_i}{\partial t} =0. 
\end{equation}
This condition means that in saddle
points the classical frequencies should be commensurable
\begin{equation}
\frac{\omega_1}{\omega_2}=\frac{N_1}{N_2}.
\end{equation}
Straightforward calculation gives
\begin{equation}
\frac{\partial \vec{\tau}}{\partial t}=-\vec{\nu}
\frac{K}{\sqrt{\omega_1^2+\omega_2^2}}
\end{equation}
where the curvature $K$ is
\begin{equation}
K=\frac{1}{\omega_1^2+\omega_2^2}
(\omega_1^2\frac{\partial^2 H}{\partial I_2^2}+
\omega_2^2\frac{\partial^2 H}{\partial I_1^2}-
2\omega_1\omega_2 \frac{\partial^2 H}{\partial I_1 \partial I_2}).
\label{108}
\end{equation}
As 
\begin{equation}
\frac{\partial^2 (N_1 I_1+N_2 I_2)}{\partial t^2}=\sum_{i=1}^2
N_i\frac{\partial \tau_i}{\partial t}=-\frac{K}{\sqrt{\omega_1^2+\omega_2^2}}
\sum_{i=1}^2N_i \nu_i,
\end{equation}
the oscillating part of the level density takes the form
\begin{equation}
d^{osc}(E)=\frac{1}{\hbar^{3/2}}\sum_{N_1,N_2\ge 0} P
\exp
(\frac{i}{\hbar}S_{cl}(E)-i\frac{\pi}{2}(N_1\mu_1+N_2\mu_2)
-i\frac{\pi}{4}\mbox{sgn}K)
+c.c.,
\label{110}
\end{equation}
where the pre-factor $P$ is
\begin{equation}
P=\frac{1}{\sqrt{(N_1\omega_1+N_2\omega_2)|K|}}.
\end{equation}
The classical action of resonant periodic tori (= periodic orbit)
\begin{equation}
S_{cl}(E)=2\pi (N_1I_1+N_2I_2),
\label{112}
\end{equation}
is defined by special values of action variables $I_i=I_i(N_1,N_2,E)$ for which the
following two equations are valid 
\begin{equation}
H(I_1,I_2)=E
\label{113}
\end{equation}
and
\begin{equation}
N_2 \frac{\partial H(I_1,I_2)}{\partial I_1}=
N_1 \frac{\partial H(I_1,I_2)}{\partial I_2}.
\label{114}
\end{equation}
Introducing the classical period of motion
\begin{equation}
T(E)=\frac{\partial S_{cl}(E)}{\partial E}=2\pi\frac{N_1}{\omega_1},
\label{115}
\end{equation}
one can rewrite the pre-factor in the following form
\begin{equation}
P=\sqrt{\frac{T(E)}{2\pi |\kappa|}},
\label{116}
\end{equation}
where
\begin{equation}
\kappa=N_1^2\frac{\partial^2 H}{\partial I_2^2}+
N_2^2\frac{\partial^2 H}{\partial I_1^2}-
2N_1N_2 \frac{\partial^2 H}{\partial I_1 \partial I_2}.
\label{117}
\end{equation}
Eq.~(\ref{110}) is the semi-classical trace formula for general integrable
systems and is a generalization of the trace formula for rectangular billiard
given by Eq.~(\ref{17}). The important difference between these two formulas
is that the periodic torus action is given by a simple formula (\ref{12}) for
the latter but only by implicit formulas (\ref{112})-(\ref{114}) for the former. 

The first step to apply the saddle point method similar to the one
discussed in previous Sections to the general trace formula (\ref{110})
is to find the expansion of the classical action
$S_{cl}(N_1+\delta N_1,N_2+\delta N_2,E)$ into series of $\delta N_i$.

By differentiating Eqs.~(\ref{112})-(\ref{114}) one obtains that this
expansion (which is a generalization of Eq.~(\ref{28})) has the following form
\begin{eqnarray}
S_{cl}(N_1+\delta N_1,N_2+\delta N_2,E)&=&
S_{cl}(N_1,N_2,E)+2\pi (I_1\delta N_1 +I_2\delta N_2)
\nonumber\\
&+&Q(N_2\delta N_1-N_1\delta N_2)^2,
\end{eqnarray}
where 
\begin{equation}
Q=\frac{2\pi^2}{T(E)\kappa},
\label{118}
\end{equation}
and all other notations are the same as above.

Taking into account that $I_1, I_2$ do not depend on the common factor of
integers $N_1,N_2$ and assuming that values of action variables $I_i$ for
different primitive periodic tori are non-commensurable one concludes that
the only possibility of existence of non-trivial saddle points (i.e. the 
cancellation of linear terms from different periodic tori in the exponent) is
exactly the same conditions as in previous Sections. For example, for two
periodic tori the saddle point manifold has the form exactly (up to
notations) the same as in Eq.~(\ref{35})
\begin{eqnarray}
&&N_1^{(1)}=r^{(1)} n_1+\delta N_1,\;
N_2^{(1)}=r^{(1)} n_2+\delta N_2,\nonumber \\
&&N_1^{(2)}=r^{(2)}  n_1+\delta N_1,\;
N_2^{(2)}=r^{(2)} n_2+\delta N_2,
\label{119}
\end{eqnarray}
where superscript denotes values for different periodic tori. As before, we
assume that integers $n_1$ and $n_2$ have no common factor. 

All further calculations can be performed exactly  as it has been done for
rectangular billiard in Section 4. The only difference with the latter case
is different expressions of
the pre-factor in the trace formula (\ref{116}),  the coefficient of the
quadratic form (\ref{118}), and the period of the motion. But it is easy to
check that
\begin{equation}
P\sqrt{\frac{\pi}{|Q|}}\frac{1}{T}=\frac{1}{2\pi},
\label{120}
\end{equation}
which agree with Eq.~(\ref{79}) and all formulas derived for the oscillating
parts of products of periodic orbit contributions for rectangular billiard
remain valid for generic integrable systems as well. Note that from 
Eq.~(\ref{46}) it follows that the total Maslov phase will have the correct
value.

To apply the saddle point method to smooth terms one need to know the
mean density of periodic orbits for general integrable systems. Though it is
possible to do it by applying the Hannay-Ozorio de Almeida method to integrable
systems \cite{4} it is instructive to compute it directly from
Eqs.~(\ref{112})-(\ref{114}). 

Let us define the density of periodic orbits with fixed period (\ref{115}) 
as 
\begin{equation}
\rho(\tau)=\int_{0}^{\infty}\delta (T(N_1,N_2,E)-\tau)dN_1dN_2.
\end{equation}
Introducing polar coordinates $N_1=R\cos \phi $,
$N_2=R\sin \phi$ one gets 
\begin{equation}
\rho(\tau)=
\int \delta (\frac{2\pi R\cos \phi }{\omega_1}-\tau)RdRd\phi=
\frac{\tau}{4\pi^2}\int (\omega_1^2+\omega_2^2) d\phi,
\end{equation}
This integral can be transformed into an integral over  classical actions $I_i$
by noting that they are functions of $\phi$ and $E$
defined by Eqs.~(\ref{113}) and (\ref{114}). Straightforward calculations give
\begin{equation}
\mbox{det} \left ( \begin{array}{lr} 
\partial I_1/\partial E & \partial I_1/\partial \phi\\
\partial I_2/\partial E & \partial I_2/\partial \phi \end{array}
\right ) =\frac{1}{K},
\end{equation}
where $K$ is the curvature defined in Eq.~(\ref{108}). As
$\omega_2/\omega_1=\tan \phi$
\begin{equation}
K=\cos^2 \phi \frac{\partial^2 H}{\partial I_2^2}+
\sin^2 \phi \frac{\partial^2 H}{\partial I_1^2}-
2\cos \phi \sin \phi  \frac{\partial^2 H}{\partial I_1 \partial I_2}.
\end{equation}
Therefore
\begin{equation}
d\phi =K\delta(E-H(I_1,I_2))dI_1dI_2,
\end{equation}
and in particular
\begin{equation}
  \rho(\tau)=\frac{\tau}{4\pi^2}\int (\omega_1^2+\omega_2^2)K
  \delta (E-H(I_1,I_2))dI_1dI_2.
\end{equation}
The computation of smooth terms can be performed exactly as above but with
different values of pre-factor, $P$ and the coefficient, $Q$,  of the 
quadratic form in the exponent. We have
\begin{equation}
P=\sqrt{\frac{\cos \phi}{ \omega_1 K}},
\label{156}
\end{equation}
and
\begin{equation}
Q=\frac{\pi \omega_1}{\cos \phi K }.
\label{157}
\end{equation}
Because the reduced period 
\begin{equation}
T=2\pi \frac{\cos \phi}{\omega_1},
\label{158}
\end{equation}
the combination of these terms in Eq.~(\ref{120}) has  the same value
as before and all integrals except
the integral over $\phi$ rest exactly as in Section \ref{smooth}. The  last 
integral takes the form
\begin{equation}
\int P\sqrt{\frac{Q}{\pi}}d\phi=\int \frac{d\phi}{K}=
\int \delta(E-H(I_1,I_2))dI_1dI_2=\bar{d},
\label{159}
\end{equation}
as it should be to reproduce the smooth part of bootstrap conditions (\ref{23}).

Though one can compute any correlation functions of classical actions with
arbitrary definition
it is convenient instead of Eq.~(\ref{200})  to define the 2-point correlation 
function of differences of actions for general integrable systems as follows
\begin{equation}
R_2(s)=\sum_{p,p'}\delta(s-S_p+S_{p'})
P_p P_{p'}e^{-i\pi (\mu_{p'}-\mu_{p})/2} a(T_p,T_{p'}),
\end{equation}
where $T_p$ are periodic orbit periods, $\mu_p$ are the Maslov indices for
the problem considered, $P_p$ are pre-factors in the
trace formula given by Eq.~(\ref{116}) (without powers of the Planck
constant), and $a(T_p,T_{p'})$ is a certain weighted function. 

Separating the diagonal and oscillatory contributions as in Section \ref{correlation}
\begin{equation}
R_2(s)=\frac{1}{2\pi}\int d\tau e^{i\tau s}(f^{(diag)}(\tau)
+f^{(osc)}(\tau)),
\end{equation}
and using Eqs.~(\ref{156})-(\ref{159})  one gets 
\begin{equation}
f^{(diag)}(\tau)=\sum_{p}P_p^2a(T_p,T_p)=\frac{1}{2\pi}\int dT a(T,T) \int
\frac{d\phi}{K}=\bar{a}\bar{d},
\end{equation}
where 
\begin{equation}
\bar{a}=\frac{1}{2\pi}\int dT a(T,T).
\end{equation}
The computation of oscillatory terms can be performed exactly as in Section
\ref{correlation}. Taking into account the relation (\ref{120}) it is easy
to check that
\begin{equation}
f^{(osc)}(\tau)= \frac{\bar{a}}{\sqrt{\tau}}\sum_{p}P_pe^{i\tau
  S_p-i\pi\mu_p/2-i\pi \mbox{\small sgn}K/4}.
\end{equation}
Therefore
\begin{equation}
f(\tau)=\hbar^2 d(E,\hbar)|_{\hbar=1/\tau},
\end{equation}
where $d(E,\hbar)$ is the level density of the quantum problem considered
which  generalizes the results in Refs.~\cite{14}, \cite{15} about duality in
integrable systems.

\section{Conclusion}

The main goal of this paper was to develop a new method which permits to
calculate sums of product of periodic orbit contributions for generic 
integrable
systems beyond the diagonal approximation. In the previous Sections it was 
demonstrated how different quantities like bootstrap conditions (\ref{23}), 
mean value of  Green function products (\ref{82}), 2-point correlation 
function of classical actions (\ref{200}), etc., can be computed for 
integrable systems directly from semi-classical trace formulas using 
this method. More applications of this method will be discussed elsewhere
\cite{17}.

The main ingredient of the method is the existence of hidden saddle points
in the summation over periodic orbits. The saddle point configurations
consist of periodic orbits whose integer vectors are mutually parallel. 
The saddle point manifold includes orbits which (i) are almost mutually parallel  
and (ii) are chosen in such a way that the total action is quadratic on
deviations from the saddle point configuration. The integration over these
quadratic forms gives the dominant contribution to the summation over
periodic orbits. By considering low-dimensional sub-manifolds it is also
possible to compute corrections to this term. 

In the calculations it was assumed that for generic integrable systems
classical actions of (almost) all  
resonant tori are non-commensurable but the method can be generalized for 
systems with geometrical symmetries and  (or) accidental degeneracy.
In Ref.~\cite{18} the method has been applied  for integrable quantum maps. 

{\bf Acknowledgments}

The author is very thankful to N. Pavloff and C. Schmit for many useful
discussions and to O. Giraud for careful reading the manuscript.


\begin{thebibliography}{99}
\bibitem{1} M.C. Gutzwiller, {\it Chaos in Classical and Quantum Mechanics},
   Springer, New York, 1990. 
\bibitem{2} M.C. Gutzwiller, J. Math. Phys. {\bf 8}, 1979 (1967); 
   {\bf 11}, 1791 (1970); {\bf 12}, 343 (1971); {\bf 18}, 806 (1977).
\bibitem{3} M.V. Berry and K.E. Mount, Phys. Prog. Phys. {\bf 35}, 315 (1972).
\bibitem{4} M.V. Berry and M. Tabor, Proc. R. Soc. London A {\bf 349}, 101
   (1976); J. Phys. A: Math. Gen. {\bf 10}, 371 (1977).
\bibitem{6} E. Bogomolny and C. Schmit, Nonlinearity {\bf 6}, 523 (1993).
\bibitem{7} M.V. Berry, Proc. R. Soc. London A {\bf 400}, 229 (1985).
\bibitem{8} E. Bogomolny and J. Keating, Nonlinearity {\bf 8}, 1115 (1995);
   {\bf 9}, 911 (1996).
\bibitem{9} E. Bogomolny, F. Leyvraz, and C. Schmit, Commun. Math. Phys.
   {\bf 174}, 577 (1996).
\bibitem{10} E. Bogomolny and J. Keating, Phys. Rev. Lett. {\bf 77}, 1472
   (1998).
\bibitem{11} J. Marloff, Commun. Math. Phys. {\bf 199}, 169 (1998).
\bibitem{12} P. Seba, Phys. Rev. Lett. {\bf 64}, 1855 (1990).
\bibitem{13} E. Bogomolny, U. Gerland, and C. Schmit, Singular Statistics
   (1997) to be published.
\bibitem{14} N. Argaman, F.M. Dites, E. Doron, J.P. Keating, A.Y. Kitaev, M.
  Sieber, and U. Smilansky, Phys. Rev. Lett. {\bf 71}, 4326 (1993).
\bibitem{15} D. Cohen, H. Primack, and U. Smilansky, Ann. of Phys. {\bf 264},
108 (1998).  
\bibitem{16} C. Schmit, private communication. 
\bibitem{17} E. Bogomolny and O. Giraud  (1999) to be published.
\bibitem{18} U. Smilansky, Trace idendities and their semiclassical
  implications, Weizmann Institute preprint, 1999.   
\end{thebibliography}
\end{document}